%% file: arxiv.tex
\documentclass[12pt,a4paper]{article}
\title{Network Thermodynamical Modelling of\\ Bioelectrical Systems: A
  Bond Graph Approach}

\usepackage{soul} 

\usepackage{xcolor}
\usepackage{framed}

\colorlet{shadecolor}{yellow}
\newenvironment{New}{}{}

\usepackage[noblocks]{authblk}
\author[1,2,4]{Peter J. Gawthrop}
\author[1,2,3]{Michael Pan}
\affil[1]{
  Systems Biology Laboratory,
  Department of Biomedical Engineering,
  Melbourne School of Engineering,
  University of Melbourne,
  Victoria 3010, Australia.
   }
   \affil[2]{Systems Biology Laboratory,
     School of Mathematics and Statistics,
     University of
      Melbourne University of Melbourne, Victoria 3010}
   \affil[3]{ARC Centre of Excellence in Convergent Bio-Nano Science and Technology, Melbourne School of Engineering, University of Melbourne, Parkville, Victoria 3010, Australia}
\affil[4]{Corresponding author
\textbf{peter.gawthrop@unimelb.edu.au}}

\usepackage{a4wide,times,siunitx}

\usepackage[super,comma,sort]{natbib}
\usepackage[hidelinks]{hyperref}
\usepackage{url}

\usepackage{siunitx}

\usepackage[section]{placeins}
\usepackage{graphicx,subfigure,fancybox,color}

\newcommand{\Si}[1]{
  (\si{#1})
  }

\include{BG-sym}
\include{BG-maths}
\include{BG-chem}
\include{BG-figs}

\begin{document}
\maketitle
\begin{abstract}
Interactions between biomolecules, electrons and protons are
essential to many fundamental processes sustaining life. It is therefore of
interest to build mathematical models of these bioelectrical
processes not only to enhance understanding but also to enable
computer models to complement \emph{in vitro} and \emph{in vivo}
experiments. Such models can never be entirely accurate; it is
nevertheless important that the models are compatible with physical
principles. Network Thermodynamics, as implemented with bond graphs,
provide one approach to creating physically compatible mathematical
models of bioelectrical systems. This is illustrated using simple
models of ion channels, redox reactions, proton pumps and
electrogenic membrane transporters thus demonstrating that the
approach can be used to build mathematical and computer models of a
wide range of bioelectrical systems.
\end{abstract}

\paragraph*{Keywords} Biological system modelling; bioelectricity;
redox reactions; network thermodynamics; computational systems biology.

\section*{Introduction}
\label{sec:introduction}
\begin{New}
In recent years, it is becoming increasingly clear that bioelectricity is involved in several key processes in cell
biology, from development to signalling to proliferation \cite{SchMelTra20,ZerAsaSoy19}.
As such, it plays an important role in many diseases and also forms
the underlying basis of several promising treatments
\cite{DjaCoo14,PchDja18,YouVedBaj20}.
Moreover, there is a strong current interest in using, or modifying, microbes
for generating energy in the form of electricity, hydrogen and
biofuels
\cite{AnwLouChe19,Gaf20,GleYinRoe20}.
Mathematical models can help to provide a mechanistic understanding of
bioelectrical systems and also to test synthetic biology constructs
\textit{in silico}\cite{PieLev16}. 
To avoid false conclusions, it is vital that such models are
constrained by physical laws such as energy conservation.  To allow
the construction of large models, a modular approach based on
interconnection of reusable components is essential.  To aid both
understanding and computer modelling, a graphical representation is
helpful.
%
The bond graph implementation of Network Thermodynamics outlined in
this paper provides a systematic way of constructing mathematical
models of living systems which are both modular and physically
constrained and have an underlying graphical representation.
\end{New}

%
Bioelectrical systems involve two physical domains: chemistry
and electricity. However, as physical domains, they are linked by the
laws of physics, and in particular by the common currency of
energy. This energy-based approach provides a unifying framework to
model bioelectrical systems.  Engineering has several ways of
modelling multi-physics energetic systems; one such method is the bond
graph methodology \cite{Pay61,GawSmi96,GawBev07,KarMarRos12}. This approach was extended some
50 years ago as an energy-based approach to modelling biomolecular
systems \cite{OstPerKat71,OstPerKat73,Per75}.
As noted by \citet{Per75}, the bond graph approach combines multi-physics modelling with graphical approaches from electrical circuit theory:
``\emph{Graphical representations similar to engineering circuit
  diagrams can be constructed for thermodynamic systems. Although
  the proverb that a picture is worth a thousand words may not be
  completely applicable, such diagrams do increase one's intuition
  about system behaviour. Moreover, as in circuit theory, one can
  algorithmically read the algebraic and differential equations that
  describe the system directly from the diagram much more easily
  than they can be constructed directly.}''.

\input{Analogies}

The Network Thermodynamic approach can be applied not
only to the \emph{chemical} properties but also to the \emph{electrical}
properties of biomolecular systems in a uniform manner.
Indeed, the bond graph approach can be seen as a formal approach to
the concept of physical analogies introduced by James Clerk
Maxwell\cite{Max71} who pointed out that analogies are central to
scientific thinking and allow mathematical results and intuition from
one domain to be transferred to another -- see Table \ref{tab:analogies}.
An issue that arises in multi-physics models is the
different dimensions and units used in each domain. For
example,
mechanical systems use force (\si{\newton}) and velocity
(\si{\metre\per\second}),
electrical systems use voltage (\si{\volt}) and current 
(\si{\ampere}) and
chemical systems use chemical potential ($\mu~\si{\joule\per\mole}$) and
molar flow ($v~\si{\mole\per\second}$).

Despite these differences, the product of the two physical covariables in each of these three
examples is energy flow \Si{\joule\per\second}. It is therefore clear that the
common quantity among all physical domains is \emph{energy}
(\si{\joule}). The commonality of energy over different physical
domains makes the bond graph approach appropriate for modelling
multi-domain systems, in particular bioelectrical
systems~\cite{GawSieKam17,Gaw17a}. Noting that the conversion factor
relating the electrical and chemical domains is \emph{Faraday's
  constant} $F\approx\SI{96485}{C.mol^{-1}}$, the \emph{chemical}
covariables $\mu$ and $v$ may be scaled by $F$ to give the pair of
covariables $\phi$ (Faraday-equivalent chemical potential) and $f$
(Faraday-equivalent flow) in \emph{electrical} units\cite{Gaw17a}:
\begin{New}
\begin{align}
  \phi &= \frac{\mu}{F}~ \Si{\volt}\label{eq:phi}\\
  \text{and } f &= F v~ \Si{\ampere}\label{eq:f} 
\end{align}
\end{New}
This reformulation is frequently used in the bioelectrical disciplines
of electrophysiology and mitochondrial energetics where it is
convenient to represent quantities in terms of voltages and currents
\cite{KeeSne09,RudSil06,NicFer13,Gaf20}.


\begin{New}
Because energy is the common quantity across physical domains, this
approach allows modular construction of large multi-physical models from
reusable components \cite{GawCra16,Hun16}. Moreover, components can be
replaced by simpler, but physically plausible, alternatives to aid
understanding and reduce computational load \cite{GawCudCra20}.
\end{New}

Equations (\ref{eq:phi})--(\ref{eq:f})  unify the description of
the electrical and chemical domains of bioelectrical systems including
the ion channel and redox examples treated in the sequel.  These
concepts are explored in the following section using electrodiffusion
and ion channels as an introductory example.  The potential of bond
graphs in representing bioelectrical systems in a modular manner is
then illustrated using a model of redox reactions and electron-driven
proton pumps in the context of the mitochondrial electron transport
chain. Finally, we consider the thermodynamics of electrogenic
transport in a model of the sodium-glucose symporter, which is
compared to experimental data.

\subsection*{Energy-based modelling -- a brief introduction}
\input{BGtable}
As noted in Table \ref{tab:analogies}, both the electrical and
chemical physical domains have energy (measured in Joules~\Si{\joule})
in common. This is the fundamental idea of energy based modelling and
means that both domains can be modelled within a common framework.
The bond graph  framework for energy-based modelling is summarised in
this section and the next section takes a closer look.
The four relevant bond graph components are listed in Table
\ref{tab:BGcomponents} and express the analogies of Table
\ref{tab:analogies}.
\begin{figure}[htbp]
  \centering
  \SubFig{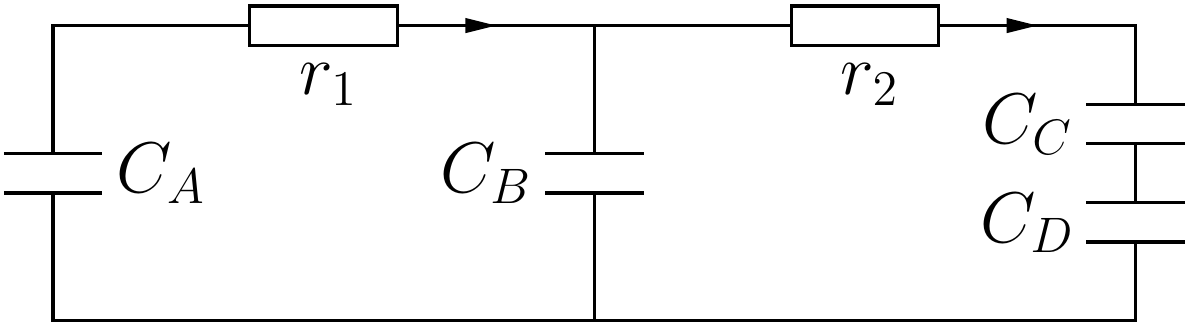}{Electrical Circuit}{0.7}
  \SubFig{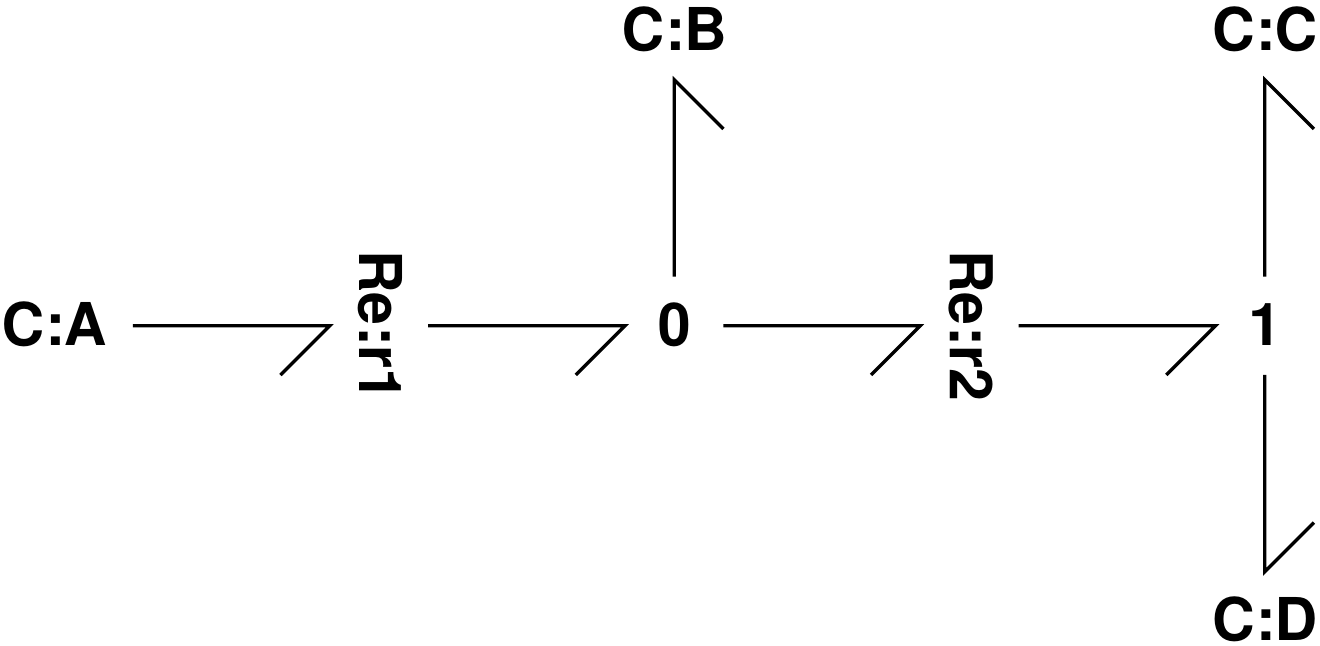}{Bond Graph}{0.7}
\begin{New}
  \caption{Analogous systems:
    (a) A reaction system \ch{A <> [ r1 ] B <> [ r2 ] C + D} comprises two reactions \ch{r1} and
    \ch{r2} and four species \ch{A}, \ch{D}, \ch{C} and \ch{D}.
    Using the analogies of Table \ref{tab:analogies}, this is analogous
    to the circuit diagram shown comprising two resistors $r_1$ and
    $r_2$ and four capacitors.
    (b) Using the components of Table \ref{tab:BGcomponents},
    both systems have the bond graph  representation shown where the
    two reactions or resistors are modelled by \BRe{r1} and \BRe{r2}
    and the 
    four species or capacitors are represented by \BC{A}, \BC{B},
    \BC{C} and \BC{D}.
    The difference between the two domains lies in the different
    equations listed in Table \ref{tab:analogies}.
    The  $\rightharpoondown$ indicate energy flows of either
    electrical energy or chemical energy and the \zero and \one
    junctions of Table \ref{tab:BGcomponents} provide the connections.
  }
  \label{fig:circuit}
\end{New}
\end{figure}
Figure \ref{fig:circuit} shows two analogous systems, a chemical
reaction system and an electrical circuit, together with the
corresponding bond graph.

\section*{Bond Graph Modelling of Electrogenic Systems}
\label{sec:bond-graph-modelling}
\begin{figure}[htbp]
  \centering
  \SubFigH{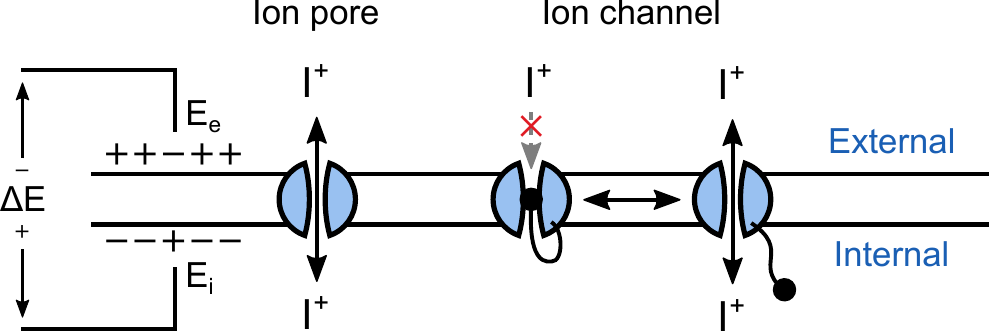}{Schematic}{0.25}
  \SubFigH{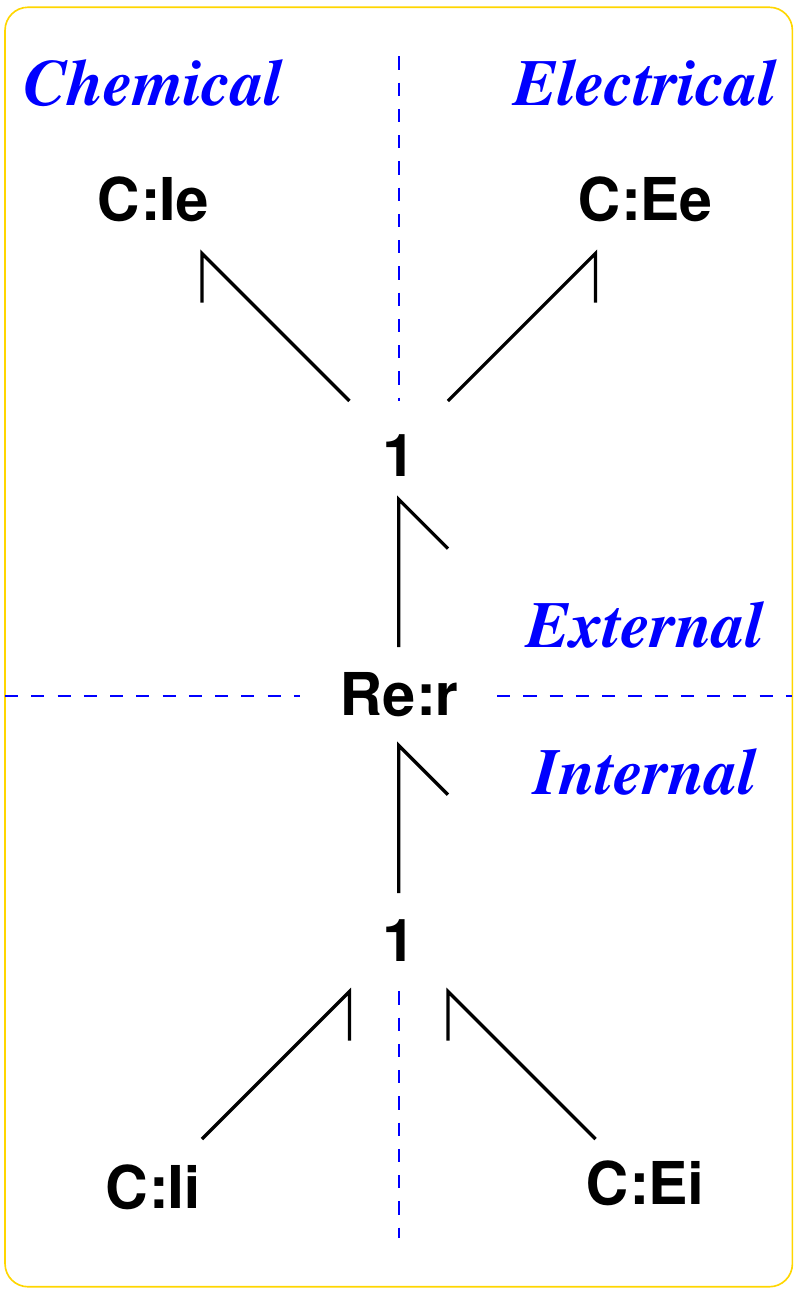}{Electrodiffusion}{0.5}
  \SubFigH{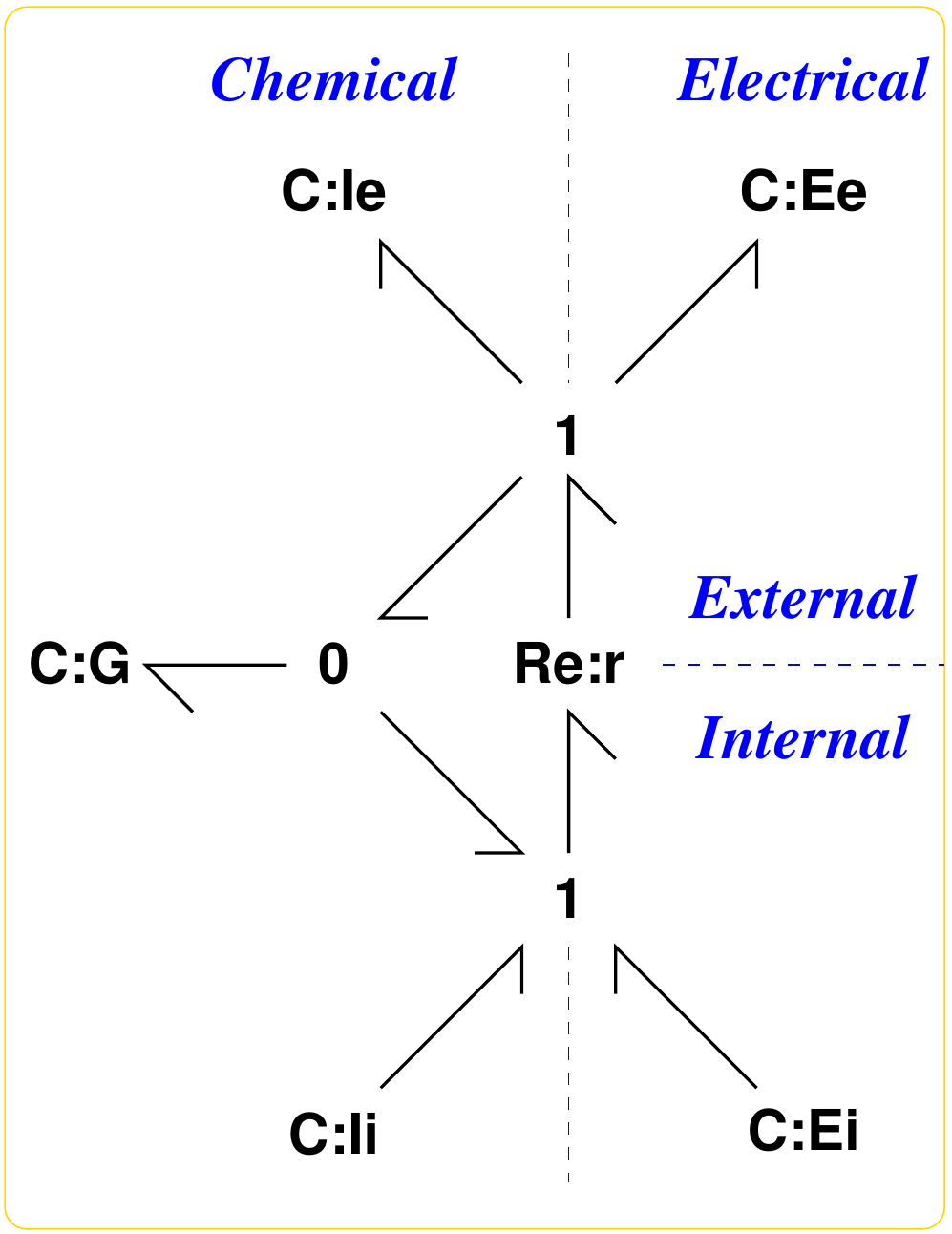}{Ion Channel}{0.5}
  \begin{New}
    \caption{
     Bond Graph Modelling of Electrogenic Systems.  %
    (a) A schematic of an ion transport through a charged membrane
    with differential voltage $\Delta E = E_i-E_e$. A
    monovalent ion \ch{I+} (which could be, for example, \ch{Na+} or
    \ch{K+}) passes though either a membrane pore (left) or an ion
    channel (middle and right). While the membrane pore is always open, the ion
    channel can switch between closed (middle) and open (right) states under the
    influence of a gating variable which may be either a ligand or a voltage.
    (b) Electrodiffusion. A monovalent ion \ch{I+} passes though a membrane
    pore where a flow from interior to exterior is regarded as
    positive.
    \BC{Ii} represents the accumulation of the ion, and \BC{Ei}
    represents the accumulation of charge, inside the
    membrane. Similarly \BC{Ie} and \BC{Ee} represent these quantities
    outside the membrane.
    \BRe{r} represents the net effect of the pores in a given area of
    the membrane.
    The \zero and \one components are \emph{junctions} connecting two
    or more bonds. The \one junction connects so that the
    \emph{flow} is common; the \zero junction connects so that the
    \emph{potential} is common.
    \C components \emph{store}, but do not \emph{dissipate}, energy;
    \Re components \emph{dissipate}, but do not \emph{store}, energy;
    \zero and \one junctions \emph{transmit}, but neither
    \emph{dissipate} nor \emph{store} energy.
    (c) Gated ion channel. The model is identical to that of (b)
    except the flow is \emph{gated} by the potential of the gate
    represented by \BC{G}. Depending on whether the channel is
    ligand-gated or voltage-gated, \BC{G} can represent either an
    accumulation of ligand or an accumulation of charge.
  }
\end{New}
\end{figure}

The bond graph representation of the network thermodynamics of
chemical systems was introduced Katchalsky and
coworkers~\cite{OstPerKat71,OstPerKat73} and  summarised by~\citet{Per75}.
In 1993, the inventor of bond graphs, Henry Paynter\index{Paynter},
said \citep{Pay93a}:
\begin{quotation}
  \emph{Katchalsky's \index{Katchalsky} breakthroughs in extending
    bond graphs to biochemistry are very much on my own mind. I remain
    convinced that BG models will play an increasingly important role
    in the upcoming century, applied to chemistry, electrochemistry
    and biochemistry, fields whose practical consequences will have a
    significance comparable to that of electronics in this
    century. This will occur both in device form, say as chemfets,
    biochips, etc, as well as in the basic sciences of biology,
    genetics, etc.}
\end{quotation}
This challenge was largely ignored until
recently when the bond graph approach was extended by
\citet{GawCra14}. 

The approach is introduced here in a tutorial fashion with reference
to a basic bioelectrical entity: electrodiffusion though a membrane
pore (Figure \ref{subfig:electrodiffusion}). A bond graph representation appears in Figure
\ref{subfig:Electrodiffusion_abg} and explained in the following
sections.

\subsection*{The energy bond}
The $\rightharpoondown$ 
symbol indicates an \emph{energy bond}:
an energetic connection between two subsystems; the
half-arrow indicates the direction corresponding to positive energy
flow. In the context of this paper, the bonds carry the Faraday-equivalent covariables $\phi$ and $f$.
These bonds connect four types of component: the four \C components
representing accumulation of chemical species (\BC{Ie},\BC{Ii}) or
electrical charge (\BC{Ee},\BC{Ei}); an
\Re component representing the membrane pore; and the
 \one junction connects the components via the bonds. These four components are now discussed in detail.

\subsection*{\C component}\label{chemical-properties}
We consider two types of \C components in this paper: those corresponding to chemical species and those corresponding to electrical charge.
The \C components representing \emph{chemical} species
generate the Faraday-equivalent potential $\phi$ of Equation
(\ref{eq:phi}) in terms of the amount of species $x$ as\cite{Gaw17a}:
\begin{align}
\phi &= \phi^\Std + \phi_N \ln \frac{x}{x^\Std}\label{eq:phi_CR}\\
\text{where }
\phi_N &= \frac{RT}{F} \approx 26 \ \si{mV}\label{eq:phi_N}
\end{align}
\begin{New}
$\phi^\Std$ is the standard potential at the standard amount
$x^\Std$. $R = 8.314 \ \si{\joule\per\kelvin\per\mol}$ is the universal gas constant and
$F$ is Faraday's
constant. The amount of species $x$ is the integral of the species Faraday-equivalent flow $f$:
\end{New}
\begin{equation}
x = \int^t f(\tau)d\tau
\end{equation}

In some cases, it is convenient to express the potential in terms of concentration $c$ as
\begin{align}
\phi &= \phi^\Std + \phi_N \ln \frac{c}{c^\Std} \label{eq:Kprime}
\end{align}
where $\phi^\Std$ is the standard potential referenced to a standard concentration $c^\Std$.

In contrast to the nonlinear chemical components, the \C components representing \emph{electrical}
capacitance are \emph{linear} and generate electrical potential $\phi$
(Volts) in terms of the charge $x=\int^t f(\tau)d\tau$ and electrical capacitance $C$ as:
\begin{align}
\phi &= \frac{x}{C}
\end{align}

\subsection*{Open systems \& Chemostats}\label{sec:open-systems-}
The concept of a \emph{chemostat}~\cite{PolEsp14} provides a way of
converting a closed system to an open system whilst retaining the
basic closed system bond graph formulation~\cite{GawCra16}.
The chemostat has a number of interpretations~\cite{GawCra18a}:
\begin{enumerate}
\item one or more species is fixed to give a constant concentration;
  this implies that an appropriate external flow is applied to balance
  the internal flow of the species.
\item an ideal feedback controller is applied to species to be fixed
  with setpoint as the fixed concentration and control signal an
  external flow.
\item as a \C component with a fixed state.
\item as a concentration clamp or fixed boundary condition.
\end{enumerate}
The chemostatted species can be thought of as external connections
turning a closed system into an open system; as such, they can be also
thought of as system \emph{ports} providing a point of interconnection
with other systems and thus leading to a \emph{modular} modelling
paradigm~\cite{Gaw17a,GawCra16,GawCurCra15}.
%

\subsection*{\Re component}
The \R component is the bond graph abstraction of an electrical
resistor. In the chemical context, a two-port \R component represents
a chemical reaction with chemical affinity (net chemical potential)
replacing voltage and molar flow replacing current
\cite{OstPerKat71,OstPerKat73}. As it is so fundamental, this two
port \R component is given a special symbol: \Re~\cite{GawCra14}.

Again, there are two versions: a non linear version corresponding to
chemical systems and a linear version corresponding to electrical systems.
In
particular, the \Re component determines a reaction flow $f$ in
terms of forward and reverse reaction potentials $\Phi^f$ and $\Phi^r$ as the
\emph{Marcelin -- de Donder} formula~\cite{Rys58}
rewritten in
Faraday-equivalent form:
\begin{align}\label{eq:MDD}
  f &= \kappa \lb \exp {\frac{\Phif}{\phi_N}}
      - \exp {\frac{\Phir}{\phi_N}} \rb
\end{align}
$\phi_N$ is given by Equation \eqref{eq:phi_N} and $\kappa$ is the
reaction rate constant.  This formulation corresponds to the
mass-action formulation; other formulations are possible within this
framework~\cite{GawCra14,GawCurCra15,GawCudCra20}.

The net reaction potential $\Phi$ is given in terms of the  forward
and reverse reaction potentials by:
\begin{equation}\label{eq:Phi}
  \Phi = \Phif - \Phir
\end{equation}
When $\Phi = 0$, the reaction is said to be at equilibrium, which also implies via Equation \eqref{eq:MDD} that the flow through the reaction $f$ is zero.

\subsection*{\zero and \one junctions}
\begin{New}
Electrical components may be connected in \emph{parallel} (where the
\emph{voltage} is common) and \emph{series} (where the \emph{current}
is common). These two concepts are generalised in the bond graph
notation as the \zero junction which implies that all impinging bonds
have the same \emph{potential} (but different flows) and the
\one junction which implies that all impinging bonds have the same
\emph{flow} (but different potentials). The direction of
positive energy transmission is determined by the bond half arrow.
As all bonds impinging on a \zero junction have the same
\emph{potential}, the half arrow implies the sign of the
\emph{flows} for each impinging bond. The reverse is true for \one
junctions, where the half arrow implies the signs of the \emph{potentials}.

These components therefore describe the \emph{network topology} of the system.
\end{New}

\subsection*{Electrodiffusion and the Nernst Potential}
\label{sec:electrodiffusion}

Figure \ref{subfig:Electrodiffusion_abg} combines these bond graph
components into a model of electrodiffusion: the flow of charged ions
though a membrane pore. The italicised text and the dashed lines are
not part of the bond graph  but add the clarity of the diagram; in
particular they emphasise the two divisions: internal/external and
chemical/electrical.

The external compartment is modelled by two \C components: \BC{Ie}
accumulating the external ion species as a chemical entity and
\BC{Ee} accumulating the external ion species as an electrical entity.
The internal compartment is modelled in a similar fashion.
The flow $f$ between the compartments is modelled by a single \Re
component; this is taken as positive if the flow is from the interior
to the exterior.
The upper \one junction ensures that the flow in to both \BC{Ie} and
\BC{Ee} is the same as that through \BRe{r}; the lower \one junction
ensures that the flow from both \BC{Ie} and \BC{Ee} is also the same as
that through \BRe{r}.

The \zero and \one junctions contribute to the network topology of the bond graph, which simultaneously (via \one junctions) distributes the flow
from reactions to species and potentials from species to reactions. This can be summarised in terms of
the \emph{stoichiometric matrix} $\NN$ as~\cite{GawCra14}:
\begin{xalignat}{2}
  \ddt{X} &= \NN f &
  \Phi = -\NN^T \phi
\end{xalignat}
where, in this case:
\begin{xalignat}{3}
\NN &=
      \left(
      \begin{matrix}1\\-1\\1\\-1\end{matrix}
  \right)&
  \XX &= \begin{pmatrix}
    x_{Ee}\\
    x_{Ei}\\
    x_{Ie}\\
    x_{Ii}\\
  \end{pmatrix}&
    \phi &= \begin{pmatrix}
    \phi_{Ee}\\
    \phi_{Ei}\\
    \phi_{Ie}\\
    \phi_{Ii}\\
\end{pmatrix}
\end{xalignat}
and $\Phi$ is the reaction potential (\ref{eq:Phi}).
In this case, $\Phi$ is given by:
\begin{equation}
  \Phi = \phi_{Ei} - \phi_{Ee} + \phi_{Ii} - \phi_{Ie}
\end{equation}
%

Defining the membrane voltage as $\Delta E = \phi_{Ei} - \phi_{Ee}$ and
using equation (\ref{eq:Kprime}):
\begin{align}
  \Phi &= \Delta E + \phi_{Ii}^\Std + \phiN \lb \ln \frac{c_i}{c_{i}^\Std} \rb - \phi_{Ie}^\Std - \phiN \lb \ln \frac{c_e}{c_{e}^\Std} \rb\\
  &= \Delta E - \phiN \ln \frac{c_e}{c_i}
\end{align}
where we have used the equalities $\phi_{Ii}^\Std = \phi_{Ie}^\Std$ and $c_{i}^\Std = c_{e}^\Std$.

The equilibrium $\Phi=0$ occurs when the membrane potential is given by:
\begin{equation}
  \Delta E = \phiN \ln \frac{c_e}{c_i}
\end{equation}
This is the well-known formula for the \emph{Nernst potential~}\cite{SteGraGil11}.


Away from equilibrium, the flow $f$ depends on the channel
characteristics and is typically modelled by the Goldman-Hodgkin-Katz
equation \cite{SteGraGil11} and can be implemented by suitably modifying
the basic \Re flow equation (\ref{eq:MDD})~\cite{GawSieKam17}.

Membrane Ion channels can be modulated (or gated) by ligands or by
voltages~\cite{Hill01}. Figure \ref{subfig:IonChannel_abg} indicates
how the electrodiffusion model can be extended by adding an additional
chemical \C component \BC{G} acting as a gate; this is directly
analogous to modulating a chemical reaction with an
enzyme~\cite{GawCra14}. The dynamics of such gates may also be
modelled using the bond graph approach~\cite{GawSieKam17,PanGawTra18}.

The interaction of two or more different ion channels sharing the same
membrane potential leads to action potentials; the following section
shows how a modular approach can be used to combine gated ion
channels.

\subsection*{Modularity: \ch{Na+} and \ch{K+} ion channels}
\begin{figure}[htbp]
  \centering
  \SubFigH{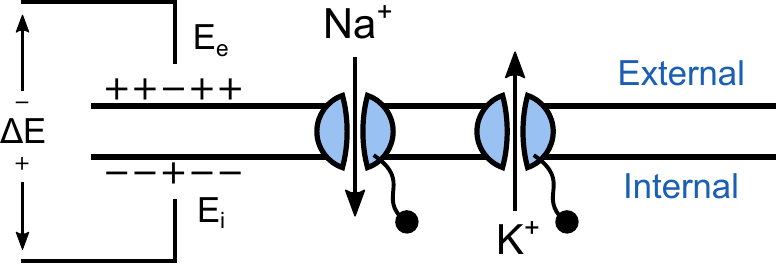}{Schematic of interacting ion channels}{0.25}
  \SubFigH{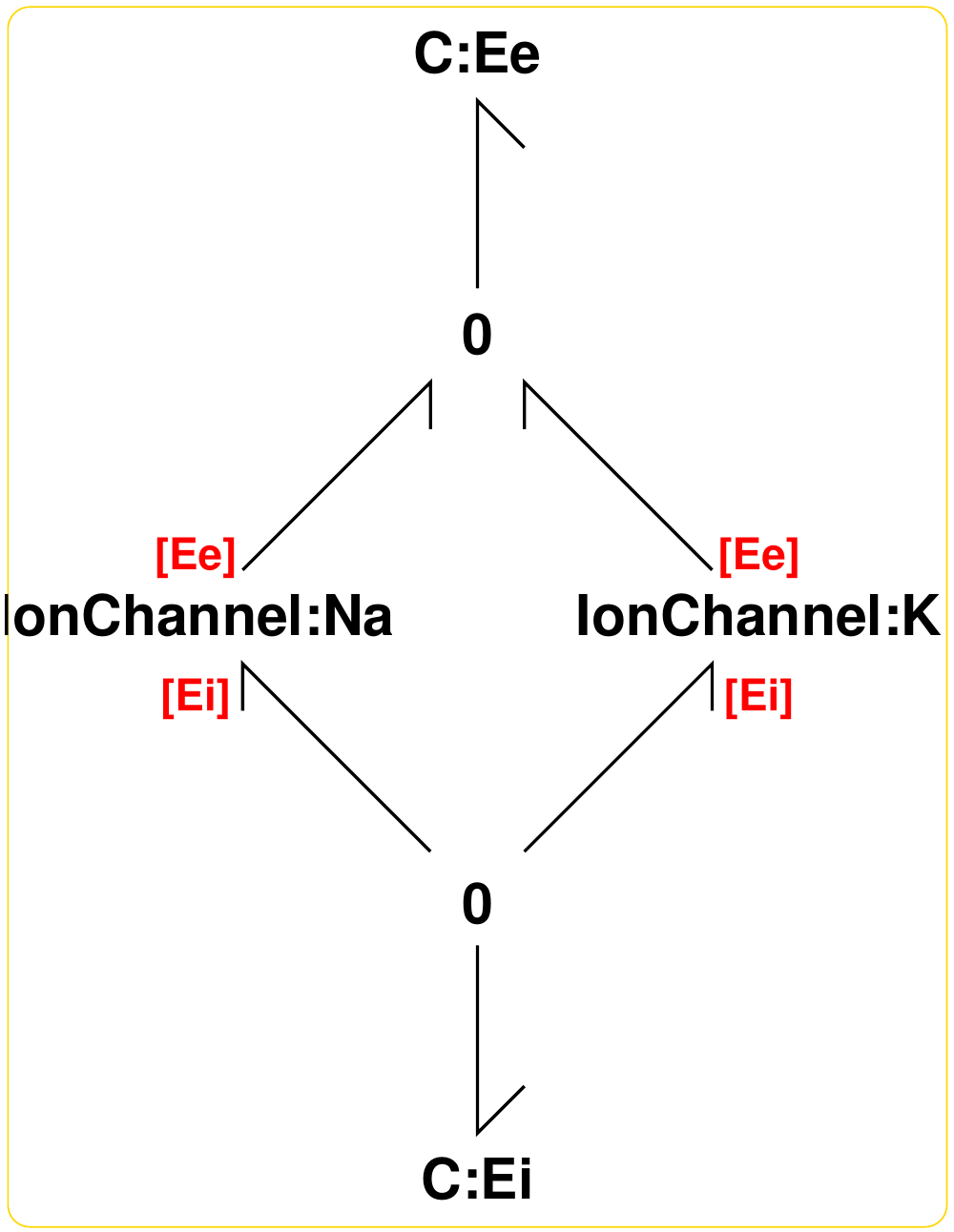}{Interacting Ion Channels}{0.4}
  \SubFigH{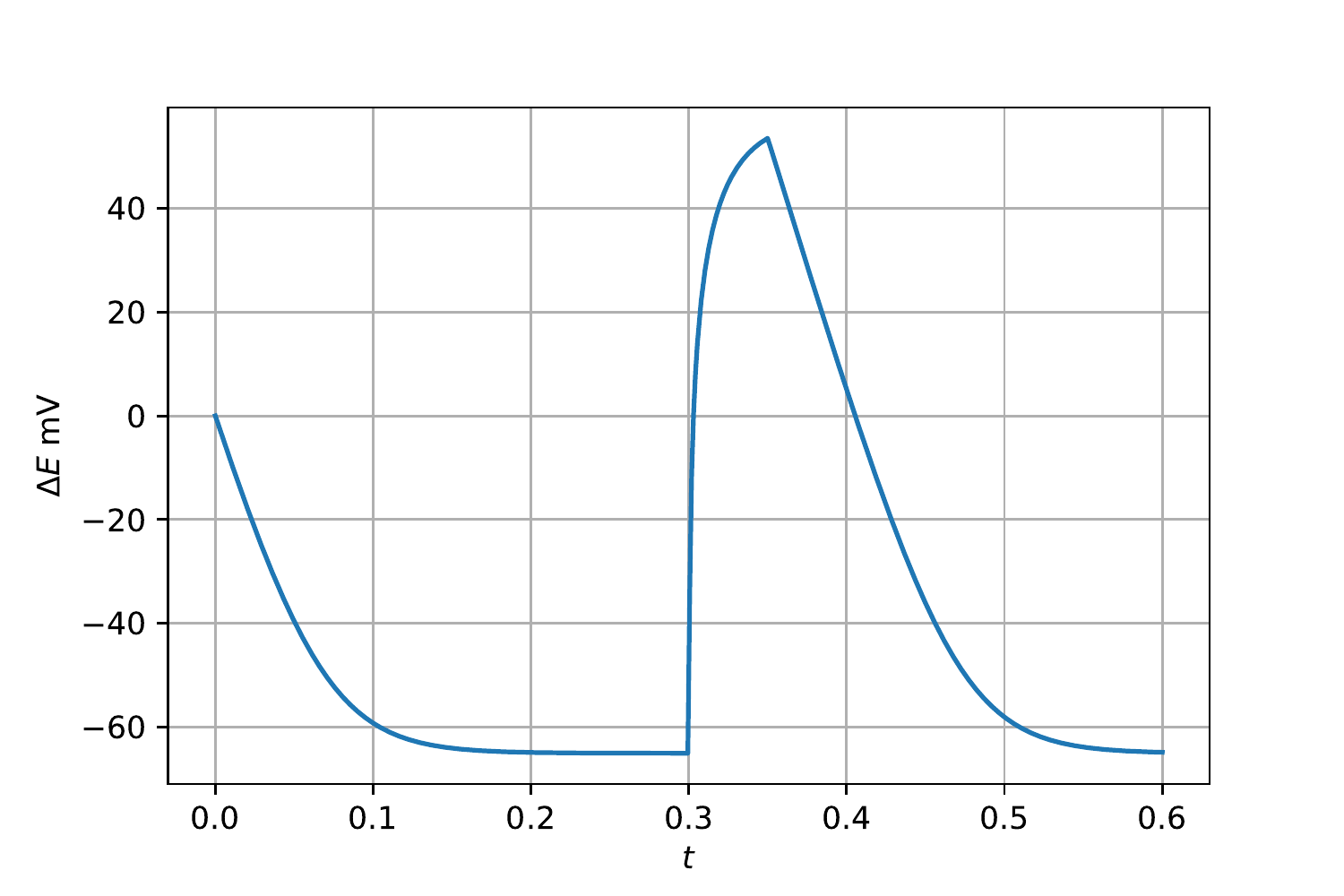}{Action potential $\Delta E$}{0.4}
  \begin{New}
  \caption{Coupled \ch{Na+} and \ch{K+} ion channels motivated by the
    Hodgkin-Huxley model of the squid giant axon\cite{HodHux52,GawSieKam17}.
    (a) A schematic of a \ch{Na+} channel and \ch{K+} channel located on the same charged membrane.
    (b) A model of this membrane can be created by coupling together two instances of the ion channel module (\BG{IonChannel}) of Figure
    \ref{subfig:IonChannel_abg}; one instance corresponds to
    the \ch{Na+} channel (\BG{IonChannel}\textbf{:Na}) and one to \ch{K+} channel (\BG{IonChannel}\textbf{:K}).
    The electrical components \BC{Ee} and \BC{Ei}, corresponding to
    the external and internal membrane
    voltages $E_e$ and $E_i$  within the ion channel
    module, become the ports \textbf{[Ee]} and \textbf{[Ei]}.
    The \zero junctions ensure that the external and internal membrane
    voltages $E_e$ and $E_i$ are shared with the corresponding ion
    channel voltages.
    (c) Action potential. Concentrations (mM) are taken from Table 2.1 of \citet{KeeSne09}:
    External. $c_{Na} = \SI{437}{mM}$, $c_{K} = \SI{20}{mM}$;
    Internal. $c_{Na} = \SI{50}{mM}$, $c_{K} = \SI{397}{mM}$
    and correspond to the model of \citet{HodHux52}.  }
\end{New}
\label{fig:modularity}
\end{figure}
A strength of the graphical nature of the bond graph approach is that individual modules can be duplicated and connected. In this example, two instances of the ion channel module (Figure \ref{subfig:IonChannel_abg}) are combined in Figure \ref{subfig:IonChannels_abg} to model the action potential (Figure \ref{subfig:action_potential}).
%
The species concentrations are encapsulated in the individual modules,
but the electrical \C components are shared via the ports. This is a
simplified version of the Hodgkin-Huxley\cite{HodHux52} model of the squid giant axon
and the corresponding \ch{Na+} and \ch{K+} concentrations are used.

In the \citet{HodHux52} model and in reality, the gating
variables are dynamically modulated by the membrane potential $\Delta E$.
However, for simplicity, we model the gating variables as
externally controlled variables in this example.
Accordingly, gating variables $G_{Na}$ and $G_{K}$ are piecewise constant functions of time:
\begin{align}
G_K &= 
\begin{cases}
10^{-6} & \text{for $0.3<t<0.35$}\\
1 & \text{otherwise}
\end{cases}\\
G_{Na} &= 
\begin{cases}
1 & \text{for $0.3<t<0.35$}\\
4.3 \times 10^{-3} & \text{otherwise}
\end{cases}
\end{align}

The time course of the membrane potential $\Delta E$ is shown in Figure \ref{subfig:action} and can be explained as follows:

$t<0.3$: $\Delta E$ moves from the initial condition of zero to
  a \emph{resting potential} of about $-65$mV.  This corresponds to
  the value in Table 2.1 of \citet{KeeSne09}; the resting potential
  depends not only on Nernst potentials of \ch{Na+} and \ch{K+} but also on the values of the
  gating potential (i.e. their relative permeability).
  
$0.3<t<0.35$: The \ch{Na+} gate opens, causing the membrane potential $\Delta E$ to move towards the Nernst potential  for
  \ch{Na+}. This results in the initial depolarisation phase of the action potential.
  
$t>0.35$ The \ch{Na+} gate closes and the \ch{K+} gate opens, causing $\Delta E$ to return to the resting potential. This completes the repolarisation phase of the action potential.

\section*{Redox Reactions and Proton Pumps}
\label{sec:exampl-redox-react}
\begin{figure}[htbp]
  \centering
  \SubFig{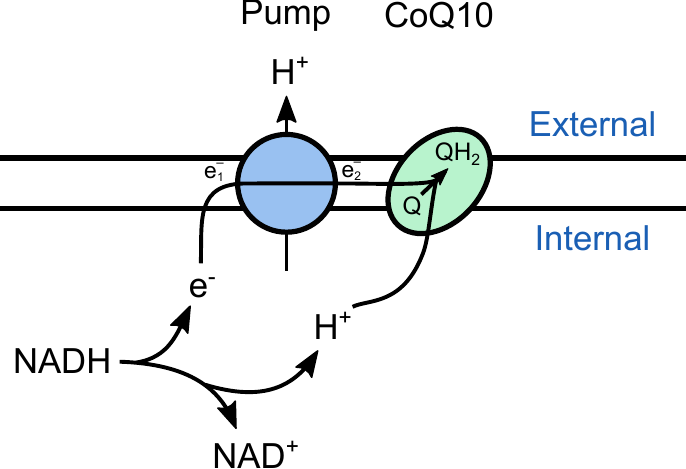}{Redox reaction schematic}{0.3}
  \SubFig{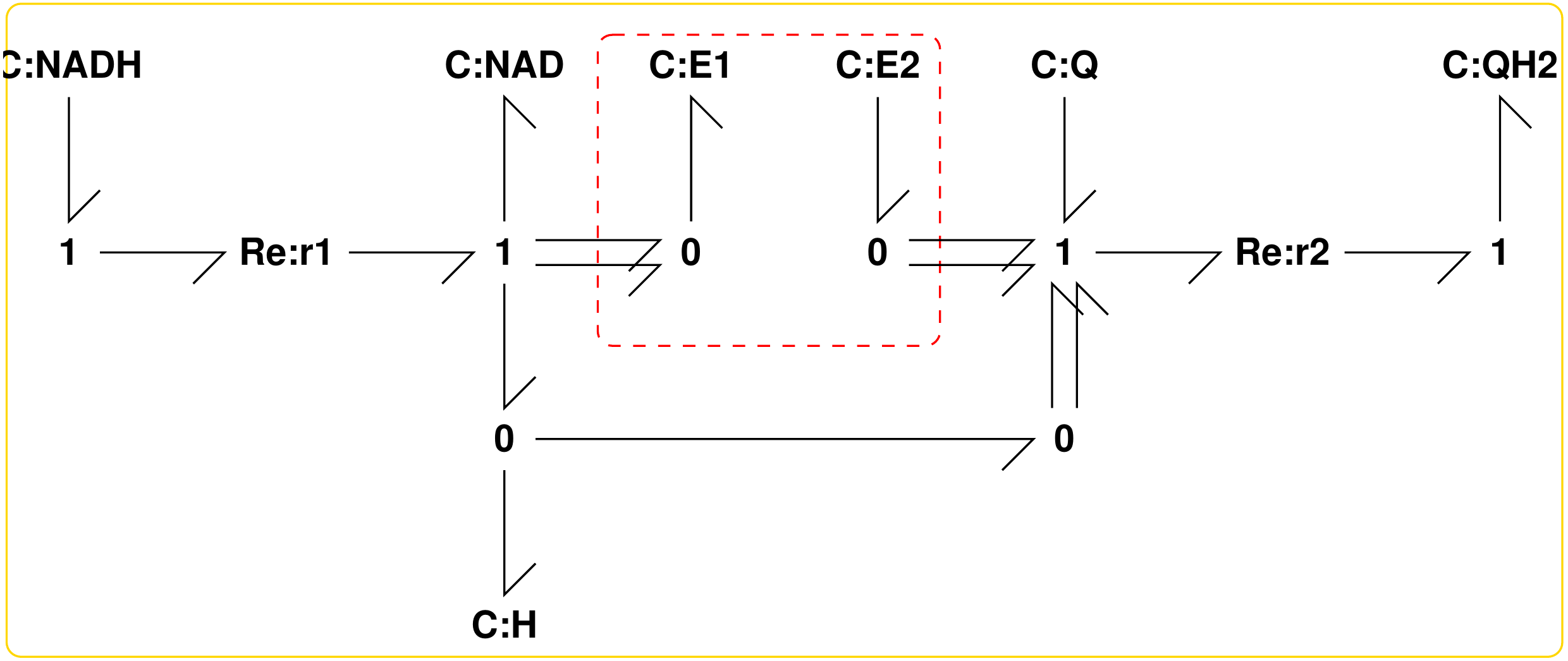}{Redox reaction}{0.6}\\
  \SubFig{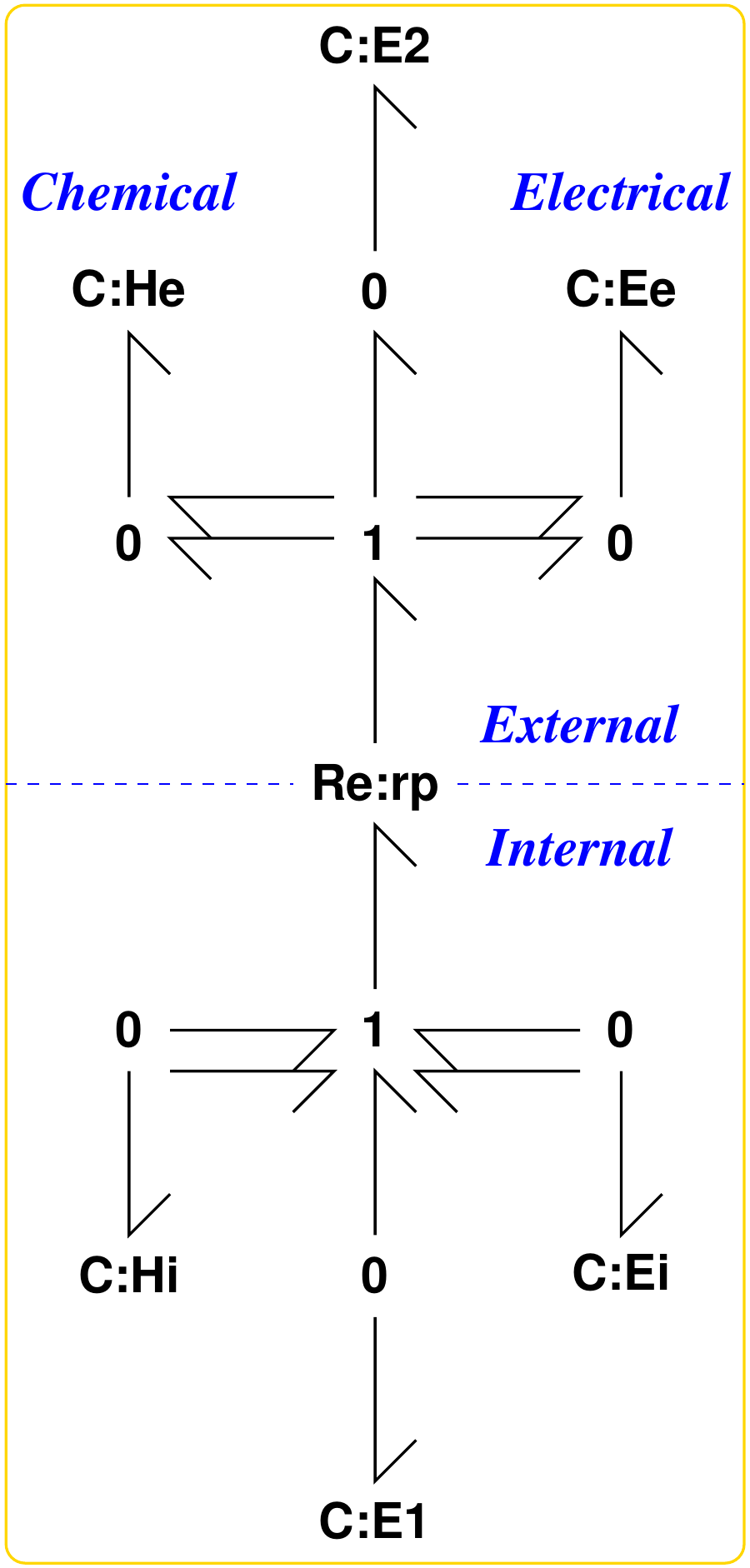}{Proton Pump}{0.2}
  \SubFig{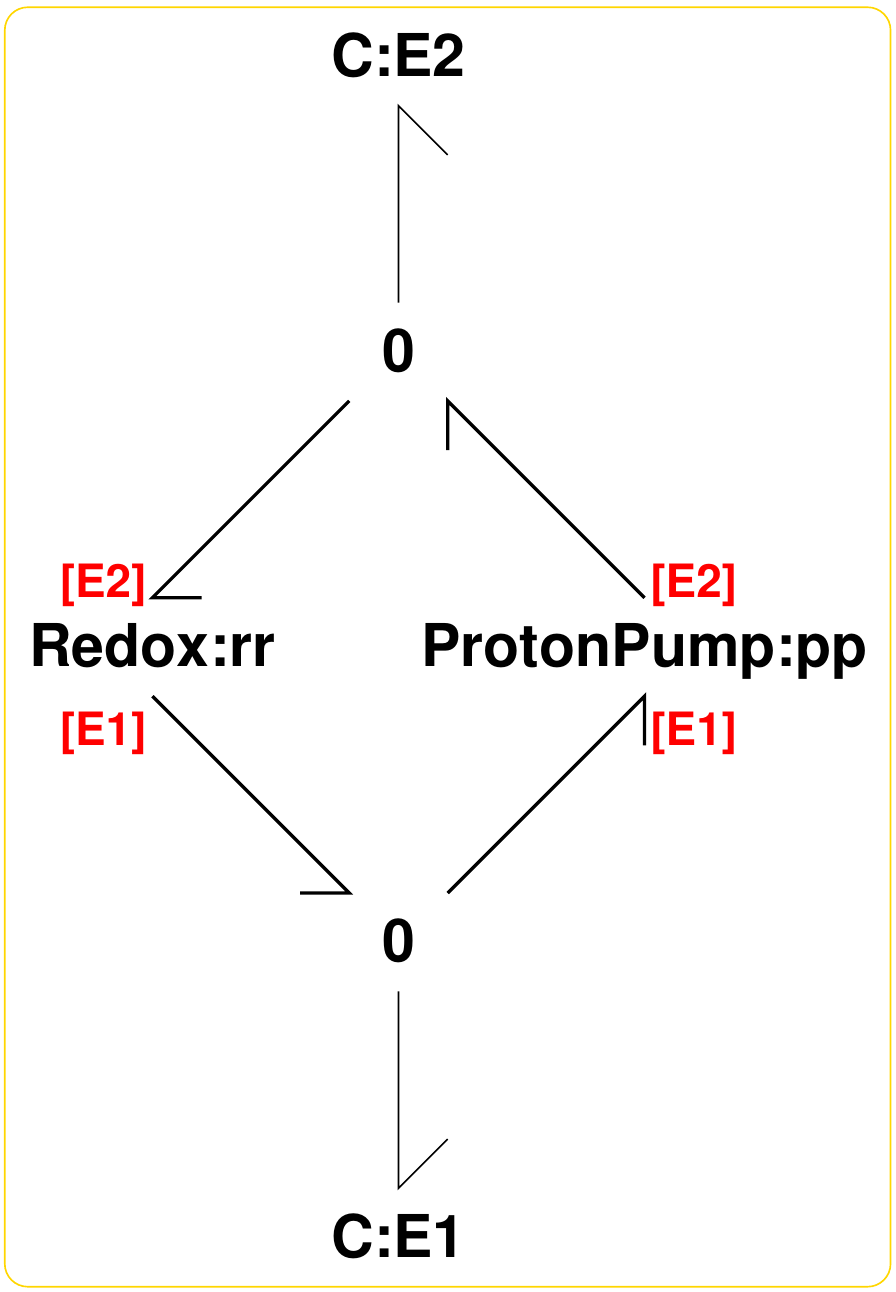}{Complex I}{0.25}
  \begin{New}
  \caption{Redox Reactions and Proton Pump. (a) A schematic of the
    redox reaction and proton pump. The electrons generated from the
    redox reaction ($\text{e}_1^-$) are shuttled through the proton
    pump to pump hydrogen into the intermembrane space. Following
    this, the electron ($\text{e}_2^-$) is then delivered to Coenzyme
    Q10. (b) A bond graph of the redox reaction.
    Half-reactions (\ref{eq:NAD_1}) and (\ref{eq:NAD_2}) are
    represented by the reaction components \BRe{r1} and \BRe{r2}
    respectively.
    The dashed box corresponds to the electrical part of the bond
    graph with \BC{E1} and \BC{E2} representing the accumulations of electrons
    \ch{e1-} and \ch{e2-}.
    The double bonds
    correspond to a stoichiometry of 2.
    (c) Bond graph of the proton
    pump. \BRe{rp} corresponds to reaction (\ref{eq:PP}).
    Internal and external protons \ch{H_i^+}and \ch{H_e^+}
    have chemical properties represented
    by \BC{Hi} and \BC{He} respectively
    and electrical  properties represented
    by \BC{Ei} and \BC{Ee} respectively.
    The redox potential difference $E_1-E_2$ drives the protons across
    the membrane.
    (d) Bond graph of the redox reaction and proton pump coupled
    together.}
\end{New}
\end{figure}
In his book \emph{Power, Sex, Suicide: Mitochondria and the Meaning of Life}\cite{Lan18a}, Nick Lane points towards the fundamental role that redox reactions play in biology, stating that
``\emph{essentially all of the energy-generating reactions of life are redox
  reactions}''.
One such energy-generating reaction that within the mitochondrial respiratory chain is
\begin{equation}\label{eq:NAD}
\ch{NADH + Q + H+ <>[ r ] NAD+ +  QH2 }
\end{equation}
As noted by \citet{NicFer13}, ``\emph{The additional facility afforded by an electrochemical
  treatment of a redox reaction is the ability to dissect the overall
  electron transfer into two half-reactions involving the donation and
  acceptance of electrons respectively}''. With this
in mind, reaction (\ref{eq:NAD}) can be divided into the reactions:
\begin{align}
\ch{NADH &<>[ r1 ] 2 e1- + H+ + NAD+ }\label{eq:NAD_1}\\
\ch{2 e2- + 2 H+ + Q &<>[ r2 ] QH2}\label{eq:NAD_2}
\end{align}
A bond graph representation of this decomposition is given in  Figure \ref{subfig:Redox_abg}.

\begin{table}[htbp]
  \centering
  \begin{tabular}{|l||l|l|l|}
\hline
Species  & $\phi^\Std~\Si{\milli\volt}$\cite{Gaw17a} & Conc. & $\phi~\Si{\milli\volt}$\\
    \hline
    \ch{NAD} & 188 & 5.02e-04\cite{ParRubXu16} & -15\\
    \ch{NADH} & 407 & 7.50e-05\cite{ParRubXu16} & 154\\
    \ch{Q} & 675 & 1.00e-02\cite{BazBeaVin16} & 552\\
    \ch{QH2} & -241 & 1.00e-02\cite{BazBeaVin16} & -365\\
    \ch{H} & 0 & 1.00e-07 & -431\\
\hline
  \end{tabular}
  \caption[Faraday-equivalent Potentials]
  {\textnormal{Faraday-equivalent potentials with $pH = 7$ and equal concentrations
    of \ch{Q} and \ch{QH2}. The potentials $\phi$ are computed from
    Equation (\ref{eq:Kprime}).
}    }
\label{tab:FEP}
\end{table}

From the bond graph, the reaction potentials of the two
half-reactions are:
\begin{align}
  \Phi_1 &= \phi_{\ch{NADH}} - \lb 2E_1 + \phi_{\ch{H}} +
           \phi_{\ch{NAD}} \rb\\
  \Phi_2 &= 2E_2 + 2 \phi_{\ch{H}} + \phi_{\ch{Q}} - \phi_{\ch{QH2}} 
\end{align}
where $E_1$ and $E_2$ are the potentials associated the electrical
capacitors \BC{E1} and \BC{E2} and the associated electrons \ch{e1-}
and \ch{e2-}.
At equilibrium $\Phi_1=\Phi_2=0$. Hence, using the values in Table
\ref{tab:FEP}, the voltages corresponding to the
capacitors \BC{E1} and \BC{E2} are
\begin{align}
  E_1 &= \frac{1}{2}\lb 154 + 15 + 431 \rb = \SI{300}{\milli\volt}\\
  E_2 &= \frac{1}{2}\lb -365 - 552 + 2 \times 431 \rb =
        \SI{-28}{\milli\volt}
\end{align}
Hence the net electronic potential $\Delta \mathcal{E} = E_1-E_2 = \SI{328}{\milli\volt}$ is available to power a proton pump (note the distinction from the membrane potential $\Delta E = E_i - E_e$ in earlier sections).
Indeed, reaction \eqref{eq:NAD} occurs within complex I of the mitochondrial
respiratory chain (Figure \ref{subfig:redox}). Complex I is a ``giant molecular proton
pump''\cite{Saz15} which uses the high-energy electron ($\ch{e1-}$) from Reaction \eqref{eq:NAD_1} to generate a proton gradient across the mitochondrial
membrane before the lower-energy electron ($\ch{e2-}$) is consumed by Reaction \eqref{eq:NAD_2}. 

In this case, two protons are pumped across the membrane for each electron.
Such an electron-driven proton pump can be modelled by the bond graph
of  Figure \ref{subfig:ProtonPump_abg}; the corresponding equation is
\begin{align}\label{eq:PP}
\ch{e1- + 2 e_i^- + 2 H_i^+ &<>[ rp ] e2- + 2 e_e^- + 2 H_e^+ }
\end{align}
The reaction affinity is 
\begin{align}
    \Phi_{rp} = \Delta \mathcal{E} - \Delta p
\end{align}
where we have defined the \emph{proton-motive force} $\pmf$ to be the sum of the
electrical and chemical potentials arising from the proton (\ch{H+})
gradient across the membrane~\cite{NicFer13}:
\begin{align}
    \Delta p = \phi_e - \phi_i - \Delta E
\end{align}
where $\phi_e$ and $ \phi_i$ are the chemical potentials of the
external and internal protons $H_e^+$ and $H_i^+$.
At equilibrium, $\Phi_{rp} = 0$ and it follows that
\begin{align}
  \pmf &= \frac{\Delta \mathcal{E}}{2} = \SI{164}{\milli\volt}\label{eq:PMF}
\end{align}
The PMF of Equation (\ref{eq:PMF}) corresponds to the equilibrium
value with standard concentrations of the species in Table
\ref{tab:FEP}. This value will change if the concentrations of the
species change or due to potential losses in the \Re
components when the flows are non zero. In fact the exchange of
species between complexes CI, CIII and CIV, in particular \ch{Q} and
\ch{QH2}, leads to an ``affinity equalisation'' of PMFs between the
three complexes with a typical PMF of approximately
$\SI{230}{\milli\volt}$~\cite{Gaw17a}.



\section*{Example: Sodium-Glucose Symporter}
\label{sec:exampl-sodi-gluc}
\begin{figure}[htbp]
  \centering
  \SubFig{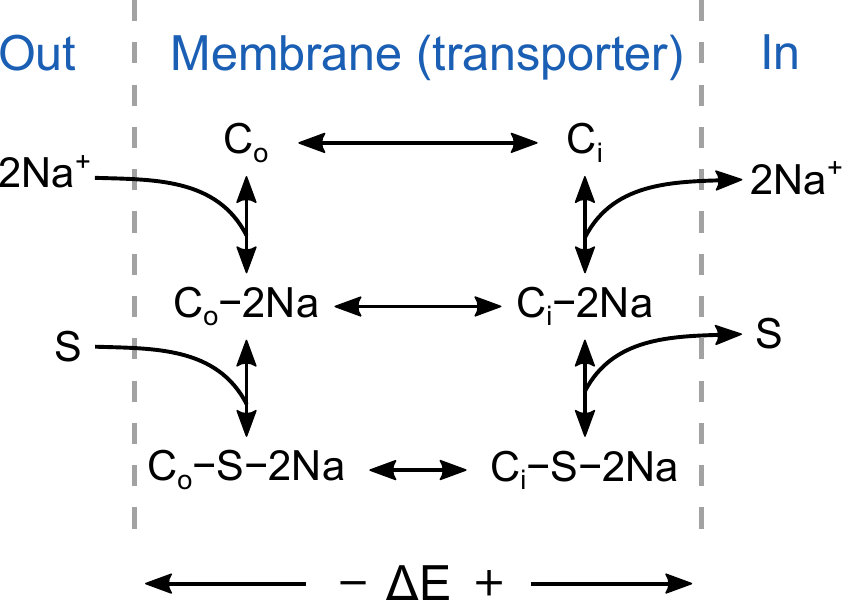}{Reaction network}{0.4}
  \SubFig{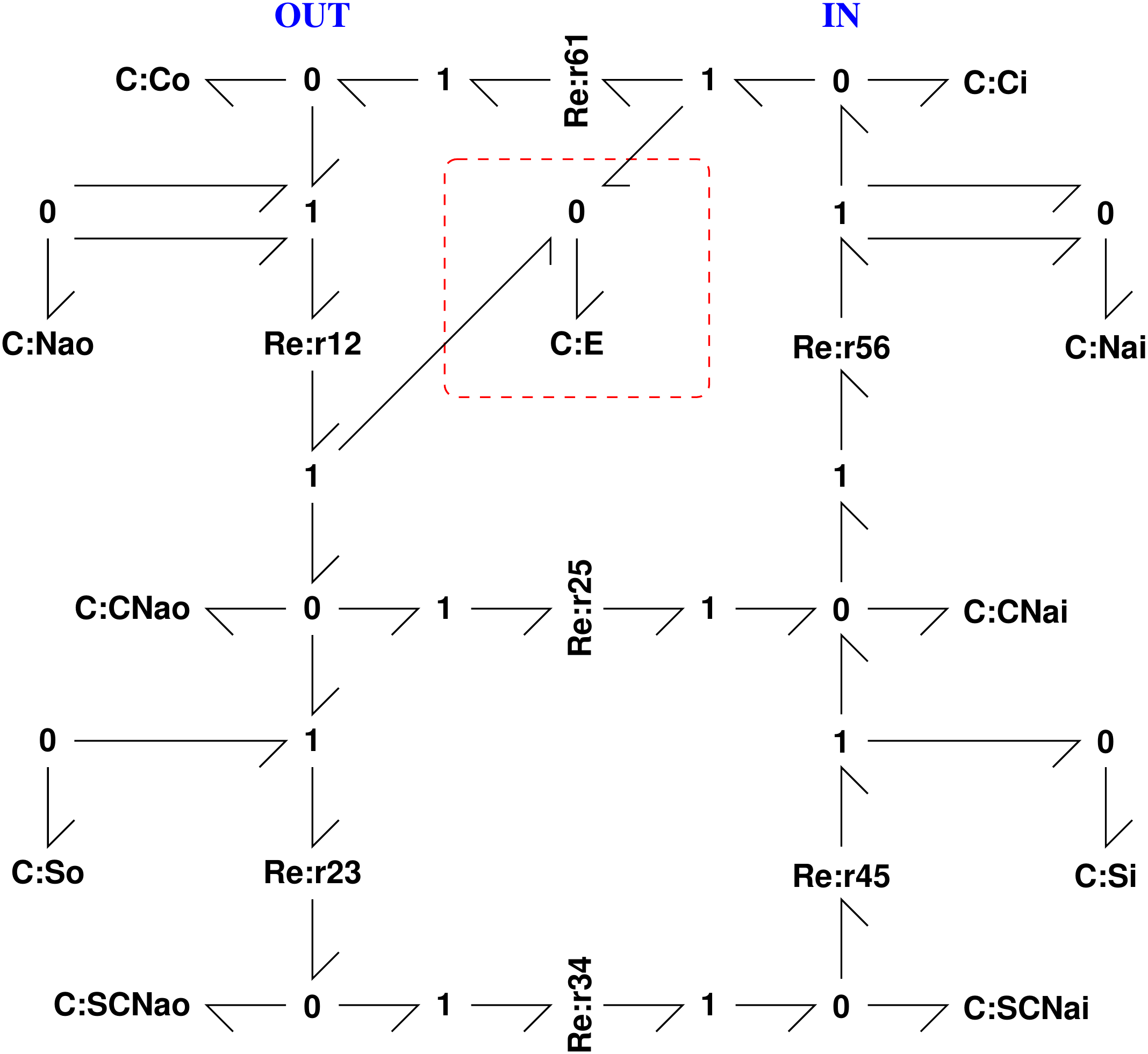}{Bond graph}{0.5}
  \begin{New}
  \caption{Sodium-Glucose Symporter. (a) A schematic of the reaction network of the transporter. Under physiological conditions, the symporter uses the natural inward electrochemical gradient of \ch{Na^+} to drive transport of the sugar glucose (S) into the cell, against a chemical gradient. The transporter operates via a six-state mechanism (outer loop), with a leakage reaction that causes \ch{Na^+} to be translocated without coupling to the transport of glucose. (b) A bond graph of the transpoter.  The cycle formed by the reactions in the outer loop (\BRe{r12}, \BRe{r23}, \BRe{r34}, \BRe{r45}, \BRe{r56} and \BRe{r61}) is driven by the outside and inside concentrations of sodium (\BC{Nao} and
  \BC{Nai} respectively) and glucose (\BC{So} and \BC{Si}). \BRe{r25} represents the leakage reaction. The contribution of the membrane potential is given by the \BC{E} component in the dashed red box.
  }
  \end{New}
  \label{fig:SGLT}
\end{figure}

\begin{figure}[htbp]
  \centering
  \Fig{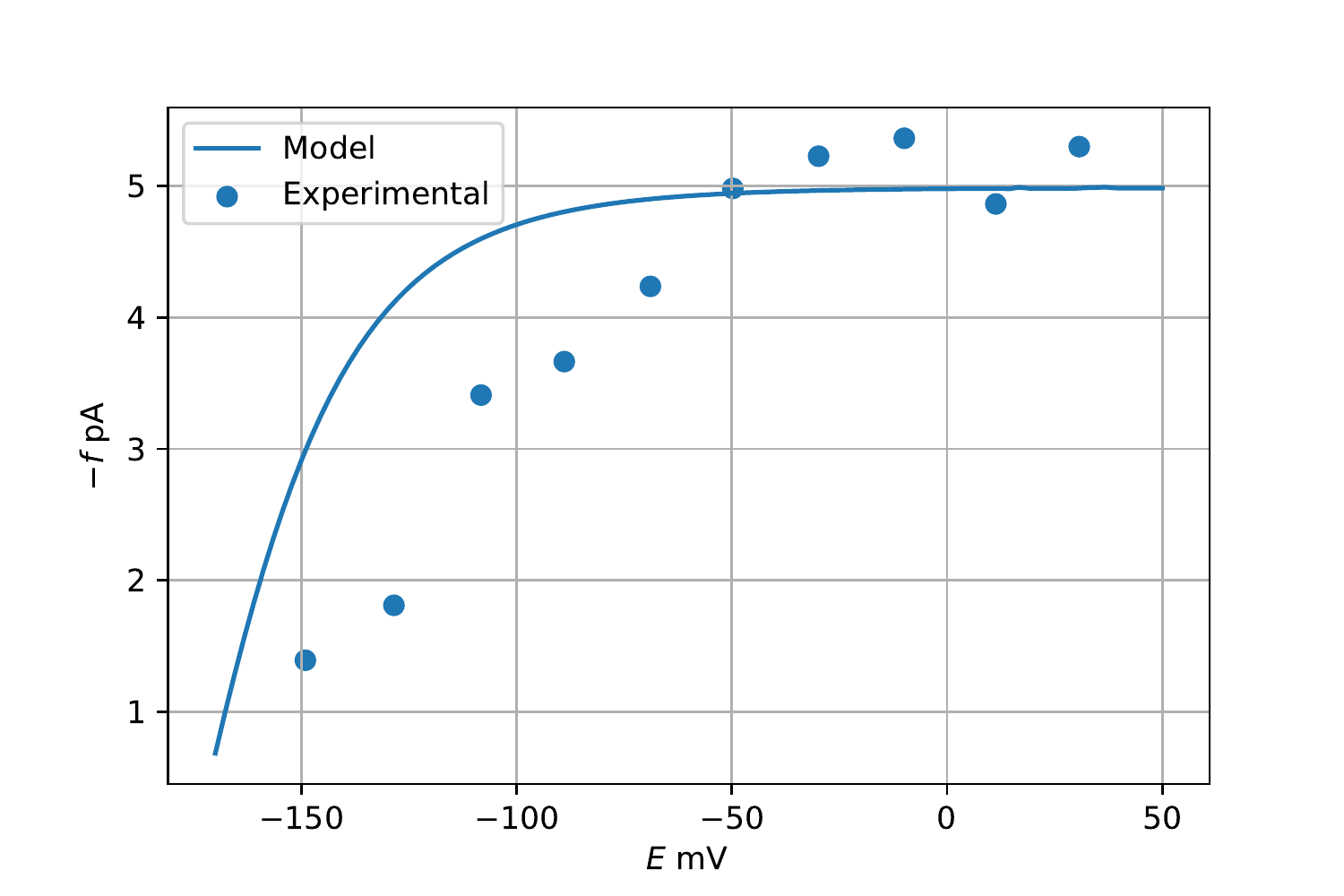}{0.7}
  \caption{Theoretical and experimental results. The experimental
    results of \citet{EskWriLoo05} are compared with the bond graph
    model with the parameters of Tables \ref{tab:parameters} and
    \ref{tab:species_parameters} and  total number of transporters $N_C=7.5\times10^7$. }
  \label{fig:sglt}
\end{figure}

The transport of substrates across a membrane is vital to cellular
homeostasis. It is widely acknowledged that the laws of thermodynamics
constrain the direction of flux through a transporter; a thermodynamic
treatment is given in \citet{Hil89} and a bond graph approaches are
developed in \citet{GawCra17} and \citet{PanGawTra19}.
In the case of electrogenic transporters that translocates a net
charge across a membrane, the effects of membrane potential must be
considered. This section looks at a particular electrogenic
transporter: the Sodium-Glucose Transport Protein 1 (SGLT1) (also
known as the \ch{Na+}/glucose symporter) which was studied
experimentally and explained by a biophysical model
\cite{,EskWriLoo05,ParSupLoo92,ParSupLoo92a,CheCoaJac95}.
%
When operating
normally, sugar is transported from the outside to the inside of the
membrane driven against a possibly adverse gradient by the
concentration gradient of \ch{Na+} (Figure \ref{subfig:transporter}).

\begin{New}
  According to \citet[p.64]{ParSupLoo92a}, ``The carrier is assumed to
bear a charge $z = -2$ with the result that the \ch{Na+} loaded carrier is
electroneutral.'' Hence, in terms of Figure \ref{subfig:ESGLT_abg}, the two
complexes corresponding to \BC{Ci} and \BC{Co} bear a charge of $-2$
whereas the loaded complexes \BC{CNao}, \BC{SCNao}, \BC{SCNai} and
\BC{CNai} are electrically neutral.

With reference to the bond graph
of Figure  \ref{subfig:ESGLT_abg}, the only reactions with electrogenic properties
are those corresponding to:
\BRe{r61} where the unloaded complex passes from the inside to
  the outside of the membrane;
\BRe{r12} where the external \ch{Na+} of \BC{CNao} binds/unbinds to
  the complex and
\BRe{r56} where the internal \ch{Na+} of \BC{CNai} binds/unbinds to
  the complex.
As discussed by \citet{ParSupLoo92a} their experimental results
suggest the ``absence of voltage dependence for internal \ch{Na+} binding,
and thus asymmetrical binding sites at the external and internal face
of the membrane.'' In the context of the bond graph model of Figure
\ref{subfig:ESGLT_abg}, this implies that the reaction corresponding to
\BRe{r56} is not electrogenic and thus electrogenic effects are
confined to the reactions corresponding to \BRe{r61} and \BRe{r12}.
These considerations give rise to Figure \ref{subfig:ESGLT_abg} where
the portion within the dashed box corresponds to electrogenic effects.
\end{New}
  A non-electrogenic version of this model is analysed in \S~1.1 of \citet{Hil89}
and in \citet{GawCra17}.

\subsection*{Parameters}
\label{sec:parameters}
\begin{table}[htbp]
  \centering
  \begin{tabular}{|l||l|l||l|l|}
    \hline
    Reac. & $k_f$ & $k_r$ & $k_{eq}=\frac{k_f}{k_r}$ & $\kappa$\\
    \hline
    \hline
    r12 & 80000 & 500 & 160 & 10.1796 \\
    r23 & 100000 & 20 & 5000 & 202.023 \\
    r34 & 50 & 50 & 1 & 505.058 \\
    r45 & 800 & 12190 & 0.0656276 & 8080.93 \\
    r56 & 10 & 4500 & 0.00222222 & 67.1184 \\
    r61 & 3 & 350 & 0.00857143 & 8.67804 \\
    r25 & 0.3 & 0.00091 & 329.67 & 0.00610777 \\
    \hline
  \end{tabular}
  \caption{\textnormal{Reaction Parameters}}\label{tab:parameters}
\end{table}

\begin{table}[htbp]
  \centering
  \begin{tabular}{|l||l|}
\hline
Species & $\phi^\Std$ (mV)\\
\hline
\hline
So & 61.71 \\
Si & 61.84 \\
Nao & 70.42 \\
Nai & 70.36 \\
Co & 98.76 \\
CNao & 104.03 \\
SCNao & -61.78 \\
Ci & -28.37 \\
CNai & -50.86 \\
SCNai & -61.78 \\
\hline
  \end{tabular}
  \caption{\textnormal{Species Parameters, referenced to a standard concentration
    of 1 M for the substrates (So, Si, Nao, Nai) and to a standard
    amount of 1 mol for transporter states (Co, CNao, SCNao, Ci, CNai,
    SCNai)}}
  \label{tab:species_parameters}
\end{table}
As is common in the literature, the experimentally derived reaction
parameters are expressed in terms of forward $k_f$ and reverse $k_r$ rate
constants whereas bond graph models are parameterised in terms of the 
standard potential $\phi^\Std$ (\ref{eq:phi_CR}) and reaction rate constant
$\kappa$ (\ref{eq:MDD}). Thus these parameters need to be converted into equivalent bond graph parameters.
The thermodynamically consistent reaction kinetic parameters listed in the first two columns of Table
\ref{tab:parameters} (given in Figure 6B of \citet{EskWriLoo05}), and are converted into ten equivalent thermodynamic constants using the methods described in \citet{GawCra17}.

The total number of transporters $N_C$ per unit area corresponding to
the experimental situation reported by \citet{EskWriLoo05} is a key
parameter.  It not only determines the steady state flows but also the
transient time constants; here it is adjusted to fit the data shown in
Figure \ref{fig:sglt}. The fitted theoretical results are compared
with experimental values in Figure \ref{fig:sglt}.

\begin{New}
  No model is definitive but rather is a step towards understanding a
  system and suggesting further experiments. For example, the model
  presented here assumes that the transmembrane current is entirely
  due to tranlocated charge. Charge leakage could be included as an
  ohmic resistance associated with the electrical \BC{E} component and
  the corresponding resistance fitted to the model.  As well as
  potentially allowing a better fit to the data, such a refined model
  would suggest further experimental verification.
\end{New}

The inhibition of the \ch{Na+}/glucose symporter has been suggested as
a treatment for type 2 \emph{diabetes mellitus}~\cite{Sch20}. It is
hoped that physically based models, such as that
developed here, will eventually lead to computational approaches to
understanding such diseases  and evaluating treatment strategies.

\section*{Conclusion}
\label{sec:conclusion}
The study of bioelectrical systems can be greatly facilitated by a modelling framework that is both graphical and physically consistent. In this perspective, a series of examples has illustrated how Network Thermodynamics, as
implemented with bond graphs, provides a graphical framework for modelling
bioelectrical systems while ensuring thermodynamic consistency. This
is made possible by treating electricity and chemistry in a unified fashion where
potentials and flows are given in the same electrical units.

The method is naturally modular and any \C component within a system
can be used as a port to connect to other subsystems. This provides a
flexible hierarchical approach to creating large models of
bioelectrical systems. Moreover, the use of energy ports allows one
particular model of a subsystem to be replaced by another with the
same ports, allowing one to swap submodels within a large model to
allow high detail in one part whilst simplifying the rest. This
approach is made possible by the fact that even low-resolution
submodels are thermodynamically consistent; for example it is possible
to build a low resolution but \emph{physically plausible} model of the
mitochondrial electron transport chain~\cite{GawCudCra20}.

Energy flow in the form of molecules, protons and electrons shapes
cell evolution~\cite{LanMar10,Lan20};
in the same vein, energy is central to synthetic biology
~\cite{ZerCheSoy18,BerUedKur19,DelCheWad20,RoeZur20}.
Hence it is appropriate to use the energy-based modelling approach to
bioelectricity to not only evaluate natural systems
from an evolutionary point of view but also to evaluate novel
bioelectrical approaches to synthetic biology such as, for example,
the generation of electricity from waste and making microbial
communities form specific patterns \cite{Bra19}.

A set of Python-based tools
\url{https://pypi.org/project/BondGraphTools/} is available to assist
model building~\cite{CudGawPanCra19X} and to provide avenues for the
automation of model development.

\section*{Acknowledgements}
  Peter Gawthrop would like to thank the Melbourne School of
  Engineering for its support via a Professorial Fellowship, and both
  authors would like to thank
  Edmund Crampin  for help, advice and encouragement, and
  Peter Cudmore for developing the BondGraphTools software package.
  This research was in part conducted and funded by the Australian Research Council Centre of Excellence in Convergent Bio-Nano Science and Technology (project number CE140100036). 
  The authors would like to thank the reviewers and editor for
  helpful comments which improved the paper.    


\input{arxiv.bbl}
\end{document}

%% file: BG-sym.tex
\newcommand{\BG}[1]{\text{\sffamily\textbf{#1}}}

\newcommand{\C}{\BG{C }}

\newcommand{\R}{\BG{R }}

\newcommand{\one}{\BG{1 }}
\newcommand{\zero}{\BG{0 }}

\renewcommand{\Re}{\BG{Re }}

\newcommand{\BGL}[2]{$\BG{#1}$:$\mathbf{#2}$} 

\newcommand{\BC}[1]{\BGL{C}{#1}}

\newcommand{\BRe}[1]{\BGL{Re}{#1}}


%% file: BG-maths.tex
\usepackage{amsmath,amssymb,amscd,amstext,mathtools,extarrows,centernot,bm}

\newcommand{\lb}{\left (}
\newcommand{\rb}{\right )}

\newcommand{\ddt}[1]{\frac{d}{dt}{#1}}

\newcommand{\NN}{N}

\newcommand{\Phif}{{\Phi^f}}
\newcommand{\Phir}{{\Phi^r}}

\newcommand{\phiN}{{\phi_N}}

\newcommand{\XX}{X}

\newcommand{\pmf}{\Delta p}

\usepackage{listings}


%% file: BG-chem.tex
\usepackage{chemformula}

\newcommand{\Std}{\ominus}



%% file: BG-figs.tex
\usepackage{graphicx,subfigure,fancybox,color}
\graphicspath{{./Figs/}}

\newcommand{\Fig}[2]{
 \includegraphics[width=#2\linewidth]{#1.pdf}
}

\newcommand{\SubFig}[3]{
 \subfigure[#2]{
   \includegraphics[width=#3\linewidth]{#1.pdf}
   \label{subfig:#1}
 }
}

\newcommand{\SubFigH}[3]{
 \subfigure[#2]{
   \includegraphics[height=#3\linewidth]{#1.pdf}
   \label{subfig:#1}
 }
}


%% file: Analogies.tex
\begin{table}[htbp]
 \centering
  \begin{tabular}{|l|l|l|}
    \hline
              &Electrical &Chemical\\
    \hline
     Potential&Voltage & Gibbs energy \\
              &$V~\si{\joule\per\coulomb}$ or \si{V}
              &$\mu~\si{\joule\per\mole}$\\
    \hline              
     Flow & Current & Molar flow\\
              &$i~\si{\coulomb\per\second}$ or \si{A}
              &$v~\si{\mole\per\second}$\\
    \hline
     Quantity& Charge & Moles\\
              &$Q~\si{\coulomb}$ where $i=\dot{Q}$
              &$q~\si{\mole}$ where $v=\dot{q}$\\
     \hline
     Capacitance \C
              &$V=\dfrac{Q}{C}$
              &$\mu = \mu^\Std + RT \ln \frac{q}{q^\Std}$\\
     \hline
     Resistance \Re
              &$i=\dfrac{V_f-V_r}{\rho}$
              &$v = \kappa \lb e^\frac{\mu_f}{RT} - e^\frac{\mu_r}{RT} \rb$\\
    \hline
  \end{tabular}
  \caption{\textnormal{Analogies and Energy-based modelling.
    In both electrical and chemical domains:
    $potential \times flow = power~\si{\joule\per\second}$.
    $\mu^\Std$ and $q^\Std$ refer to standard conditions, $RT$ is the
    product of the gas constant and temperature and subscripts $f$ and
    $r$ abbreviate forward and reverse.}
  }
  \label{tab:analogies}
\end{table}


%% file: BGtable.tex
\begin{table}[htbp]
  \centering
  \begin{tabular}{|l|l|l|}
    \hline
     Component &Electrical &Chemical\\
    \hline
    \C & capacitor & species \\
    \Re & resistor & reaction \\
    \zero & parallel connection & common potential\\
    \one & seriesl connection & common flow\\
    \hline
  \end{tabular}
  \caption{\textnormal{Bond graph components express the analogies of Table
    }\ref{tab:analogies}.}
  \label{tab:BGcomponents}
\end{table}